\documentclass[runningheads]{llncs}
\usepackage[utf8]{inputenc}
\usepackage{hyperref}
\usepackage{microtype}
\usepackage{graphicx}
\graphicspath{{figs/}}

\usepackage[noend]{algorithm,algorithmic}

\usepackage{amsmath, amssymb,bm}
\usepackage{xcolor}
\usepackage{cleveref}
\usepackage[symbol]{footmisc}

\usepackage{footnote}
\makesavenoteenv{algorithm}

\newtheorem{othertheorem}{othertheorem}
\newtheorem{rem}[othertheorem]{Remark}

\newcommand{\D}{\mathbf{D}}
\newcommand{\A}{\mathbf{A}}
\newcommand{\y}{\mathbf{y}}
\newcommand{\X}{\mathbf{X}}
\newcommand{\T}{\mathbf{T}}
\newcommand{\Z}{\mathbf{Z}}

\newcommand{\bet}{\bm\beta}

\newcommand{\R}{\mathbb{R}}
\newcommand{\E}{\mathbb{E}}
\renewcommand{\S}{\mathcal{S}}

\usepackage{array}
\newcommand{\PreserveBackslash}[1]{\let\temp=\\#1\let\\=\temp}
\newcolumntype{C}[1]{>{\PreserveBackslash\centering}p{#1}}
\newcolumntype{R}[1]{>{\PreserveBackslash\raggedleft}p{#1}}
\newcolumntype{L}[1]{>{\PreserveBackslash\raggedright}p{#1}}

\definecolor{amber}{rgb}{1.0, 0.49, 0.0}

\makeatletter
\newcommand{\printfnsymbol}[1]{%
  \textsuperscript{\@fnsymbol{#1}}%
}
\makeatother

\begin{document}

\title{MISNN: Multiple Imputation via Semi-parametric Neural Networks}

\author{Zhiqi Bu\thanks{Equal contribution}
\and Zongyu Dai$^*$  \and Yiliang Zhang$^*$ \and Qi Long \\
\email{\{zbu,daizy,zylthu14\}@sas.upenn.edu, qlong@pennmedicine.upenn.edu}}


\institute{}


%
\maketitle              
%


\begin{abstract}
Multiple imputation (MI) has been widely applied to missing value problems in biomedical, social and econometric research, in order to avoid improper inference in the downstream data analysis. In the presence of high-dimensional data, imputation models that include feature selection, especially $\ell_1$ regularized regression (such as Lasso, adaptive Lasso, and Elastic Net), are common choices to prevent the model from underdetermination. However, conducting MI with feature selection is difficult: existing methods are often computationally inefficient and poor in performance. We propose MISNN, a novel and efficient algorithm that incorporates feature selection for MI. Leveraging the approximation power of neural networks, MISNN is a general and flexible framework, compatible with any feature selection method, any neural network architecture, high/low-dimensional data and general missing patterns. Through empirical experiments, MISNN has demonstrated great advantages over state-of-the-art imputation methods (e.g. Bayesian Lasso and matrix completion), in terms of imputation accuracy, statistical consistency and computation speed.

\end{abstract}

\vspace{-1cm}
\section{Introduction}

\vspace{-0.2cm}
\subsection{Missing Value Mechanisms and Imputation}
\vspace{-0.2cm}

Missing data are commonly encountered in data analyses. It is well-known that inadequate handling of missing data can lead to biased findings, improper statistical inference \cite{deng2016multiple,zhao2016multiple} and poor prediction performance. One of the effective remedies is missing data imputation. Existing imputation methods can be mainly classified as single imputation (SI) and multiple imputation (MI) \cite{rubin2004multiple}. The former imputes missing values only once while the latter generates imputation values multiple times from some distribution. In fields such as finance and medical research, linear models are often preferred as it is important to not only predict accurately but also explain the uncertainty of the prediction and the effect of features. In the interest of statistical inference, MI methods, including MISNN proposed in this paper, are more suitable as they adequately account for imputation uncertainty and provide proper inference.


In general, performances of imputation are highly related to the mechanisms that generate missing values, which can be categorized into three types: missing completely at random (MCAR), missing at random (MAR) and missing not at random (MNAR). Missing data are said to be MCAR if the probability of being missing is the same for all entries; MAR means that the missing probability only depends on the observed values; MNAR means that the missing probability depends on the unobserved missing values. Intuitively, imputation is easier under MCAR mechanisms as the missing probability is only a (unknown) constant, and therefore most methods are designed to work under MCAR. However, MAR and MNAR are usually more difficult and fewer methods perform well on these problems.




\vspace{-0.3cm}
\subsection{Feature Selection in Imputation Models}
\vspace{-0.15cm}
In many applications including gene expression and financial time series research, we need to analyze high dimensional data with number of features being much larger than number of samples. In such cases, multiple imputation, which estimates the (conditional) distribution of missing data, can be inaccurate due to the overwhelming amount of features. Existing works \cite{deng2016multiple,zhao2016multiple} propose to use regularized linear model for feature selection, before building the imputation model. Some representative models include Lasso \cite{tibshirani1996regression}, SLOPE \cite{bogdan2015slope}, Elastic Net \cite{zou2005regularization}, Adaptive Lasso \cite{zou2006adaptive}, Sparse Group Lasso \cite{friedman2010note,simon2013sparse}, etc. 

While the regularized linear models successfully reduces the number of features, they often fail to capture the true distribution of missing data due to the linear dependence on the selected features and information loss in the unselected features when building the imputation model. Hence, the corresponding inference can be significantly biased. MISNN proposed in this paper overcomes the shortcome via semi-parametric neural networks. At a high level, MISNN is a semi-parametric model based on neural networks, which divides predictors into two sets: the first set are used to build a linear model and the other is used to build neural networks, which are often regarded as non-parametric models. We highlight that the outperformance of MISNN is contributed both by its neural network and linear parts. The neural networks effectively capture the non-linear relationship in the imputation model, and the linear model, in addition to capturing the linear relationships, allows efficient MI, through maximum likelihood estimation for the regression parameters.

\vspace{-0.3cm}
\subsection{Our Contribution}
\vspace{-0.17cm}
This paper makes two contributions. Firstly, we propose MISNN, a novel imputation method that outperforms state-of-the-art imputation methods in terms of imputation accuracy, statistical consistency, and computation speed. MISNN is easy to tune, interpretable, and robust to high missing rates and high-dimensional features. Secondly, MISNN is a flexible imputation framework that can be used with any appropriate feature selection method, such as Lasso and forward-selection. Additionally, MISNN is compatible with any neural network, including under or over-parameterized networks, CNN, ResNet, dropout, and more. 

\vspace{-0.6cm}
\begin{figure}[!htb]
  \hspace*{1.5cm}
  \includegraphics[width=11cm,height=12cm]{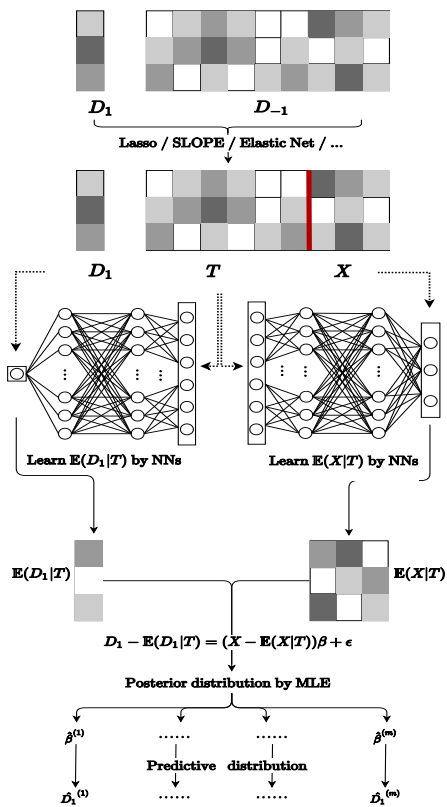}
  \vspace{-1.8cm}
  \caption{MISNN framework.}
  \vspace{-0.6cm}
\end{figure}

\vspace{0.32cm}
\section{Related Work}
\vspace{-0.1cm}

Regarding missing data imputation, SI methods have long history before the concept of MI \cite{rubin2004multiple}, of which one representative approach is the mean imputation. Recent work in SI include matrix completion approaches that translate the imputation into an optimization problem. Existing methods such as SoftImpute \cite{mazumder2010spectral} and MMMF (Maximum-Margin Matrix Factorization) \cite{srebro2004maximum} provably work under MCAR mechanisms. Meanwhile, an increasing number of MI methods are studied: MICE \cite{van2007multiple,buuren2010mice} imputes missing values through the chained equations; MissForest \cite{stekhoven2012missforest} imputes missing values via bootstrap aggregation of multiple trees. Deep generative models \cite{yoon2018gain,gondara2018mida,mattei2018miwae,li2019misgan,dai2021multiple}, including Generative Adversarial Impu-tation Nets (GAIN), are also proposed for imputation. We remark that most of the existing methods only provably work under MCAR (though some methods empirically work well under MAR).


Regularized linear models have been proposed for MI in high-dimensional data. Bayesian Lasso \cite{park2008bayesian,hans2009bayesian} estimates the posterior distribution of coefficients, while alternative approaches \cite{zhao2016multiple,deng2016multiple} de-bias the estimator from the regularized linear regression. Namely, the direct use of regularized regression (DURR) and the indirect use of regularized regression (IURR). However, linear imputation models fail to capture the potential non-linear relations in the conditional distribution of missing data. MISNN falls into this line of research, is computationally more efficient than Bayesian Lasso, and captures non-linear relations during imputation.


Recent work has highlighted the importance of trustworthiness in missing data imputation, with privacy-preserving \cite{jagannathan2008privacy,das2022imputation,clifton2022differentially} and fairness-aware \cite{martinez2019fairness,zhang2021fairness,caton2022impact,zhang2022fairness} imputation models drawing attention. MISNN has strong interpretability, allowing for better understanding of the imputation process and greater trust in the results.

\vspace{-0.2cm}
\section{Data Setup}
\vspace{-0.2cm}

Denote the data matrix by $\D\in\R^{n\times p}$, where $n$ is the number of samples/cases and $p$ is the number of features/variables. We define the $j$-th feature by $\D_j$ and its complement features by $\D_{-j}:=\D_{2:p}$ for $j\in [p]$. In the presence of missing data, $\D$ can be separated into two submatrices $\D_{\text{cc}}$ and $\D_{\text{ic}}$, where $\D_{\text{cc}}$ denotes all complete cases (i.e. all features are observed) and $\D_{\text{ic}}$ denotes all incomplete cases. We let $\D_{\text{cc},j}$ and $\D_{\text{ic},j}$ denote the $j$-th feature of complete cases and incomplete cases, respectively. We also define $\D_{\text{miss}}$, the set of missing features in $\D$, and $\D_{\text{obs}}$, the set of observed features for samples in $\D$. Briefly speaking, to impute the missing values, we fit an imputation model $g$ using $\D_{\text{obs}}$, and use $\D_{\text{ic},\text{obs}}$ as input to give imputation result $\hat{\D}_{\text{miss}}$. For the ease of presentation, we start with a single feature missing, in which only the first column in $\D$ (i.e., $\D_1$) contains missing values. We then move on to the general missing pattern with multiple features missing in \Cref{sec:experiment}. 

\vspace{-0.4cm}
\subsection{A Framework for Multiple Imputation}
\vspace{-0.2cm}
Here we provide a brief discussion about a general framework for multiple imputation, which is also adopted in MISNN. Under the above data setting, MI methods estimate the conditional distribution $\rho(\D_{\text{miss}}|\D_{\text{obs}})$ and sample imputed values from it multiple times. Assuming the distribution of $\D$ is characterized by unknown parameters $\bm\xi$, then
\vspace{-0.2cm}
\begin{align*}
\rho(\D_{\text{miss}}|\D_{\text{obs}})=\int\rho_{\text{miss}}(\D_{\text{miss}}|\D_{\text{obs}},\bm\xi)\rho_2(\bm\xi|\D_{\text{obs}})d\bm\xi
\vspace{-0.2cm}
\end{align*}
in which $\rho,\rho_1,\rho_2$ are three conditional distributions. For the $m$-th imputation, we randomly sample $\bm\xi^{(m)}$ from the posterior distribution of $\bm\xi$, i.e. $\rho_2(\bm\xi|\D_{\text{obs}})$; we then generate the $m$-th imputed data $\D_{\text{miss}}^{(m)}$ from the predictive distribution $\rho_1(\D_{\text{miss}}|\D_{\text{obs}},\bm\xi)$. With multiple imputed datasets, further analysis and inference can be conducted with the help of Rubin's rule \cite{little2002statistical,rubin2004multiple}. A detailed introduction of Rubin's rule is provided in Appendix \ref{rubin's rule}.

\vspace{-0.2cm}
\section{Multiple Imputation with Semi-parametric Neural Network (MISNN)}

At the high level, MISNN imputes the missing data in each column through a partial linear model (PLM), which takes the form
\small
$$
\hat{\D}_{1} = \X\hat{\beta} + \hat{f}(\T)
$$
\normalsize
where $(\X,\T)$, determined through feature selection, is a partition of the rest $p-1$ columns. While the choice of $\hat{\beta}$ and $\hat{f}(\T)$ can be determined in an arbitrary manner, we adopt a partialling out approach \cite{robinson1988root} (also known as the orthogonalization in \cite{chernozhukov2018double}) that can provide consistent parameter estimation if the true model takes the form ${\D}_{1} = \X{\beta} + {f}(\T) +\bm\epsilon$. To do so, we take the conditional expectation on $\T$, assuming $\E(\bm\epsilon|\T)=0$:
\small
\begin{align}
\begin{split}
\D_1&=\X\bm\beta+f(\T)+\bm\epsilon
\\
\E(\D_1|\T)&=\E(\X|\T)\bm\beta+f(\T)
\\
\D_1-\E(\D_1|\T)&=\left(\X-\E(\X|\T)\right)\bm\beta+\bm\epsilon
\end{split}
\label{eq:PLM linear equation}
\end{align}
\normalsize


Let $\mathcal{S}$ denote the set of features selected. Notice that $\T:=\D_{-1} \setminus \D_\mathcal{S}$ is explicitly removed in the last equation. Therefore, if the number of selected features can be controlled (i.e., $|\mathcal{S}|$ is small), we are left with a low-dimensional linear model (as $\X-\E(\X|\T)\in\R^{n\times |\S|}$), as long as we can estimate the mapping $\E(\D_1|\T)$ and $\E(\X|\T)$ properly. To realize the above approach, MISNN algorithm takes three key steps:
\begin{itemize}
    \item \textbf{Feature Selection: } During imputation of each missing feature, MISNN conducts feature selection to select at most $n$ features. The selected features $\X$ are expected to have significant linear correlation with the missing feature, which later will be fitted in a linear model (e.g., least squares).
    \item \textbf{Fitting Partially Linear Model: } Suppose the remaining features after the selection are denoted by $\T$, MISNN fits two neural networks to learn $\E(\D_{\text{miss}}|\T)$ and $\E(\X|\T)$, so as to derive a low-dimensional ordinary linear model \eqref{eq:PLM linear equation};
    \item \textbf{Multiple Imputation: } MISNN uses maximum likelihood to estimate parameters in \eqref{eq:PLM linear equation}, then draw $M$ times from the posterior distribution of $\hat{\bm\beta}$ and further draw $\widehat\D_{\text{miss}}$ from the predictive distribution.
\end{itemize}

Note that the first two steps in combination is closely related to DebiNet \cite{xu2020debinet}, though we do not refine ourselves to over-parameterized neural network, and we utilize two neural networks to learn $\left(\E(\D_{\text{miss}}|\T),\E(\X|\T)\right)$. In the following, we introduce MISNN in \Cref{alg: misnn} and validate the procedure of MISNN rigorously. Here we assume the missing feature is continuous. For non-continuous features, some modifications to the algorithm should be made. See details in \Cref{sec: gplm}. 

\begin{rem}
If one only focuses on the prediction, not the inference, single imputation can be conducted in \Cref{alg: misnn}. In particular, OLS can solve the linear model in step (4) and we impute by $$\widehat\D_{\text{ic},1}= \left(\X_{\text{cc}}-\E(\X_{\text{cc}}|\T_{\text{cc}})\right)\bm{\hat\beta}+\E(\D_{\text{cc}, 1}|\T_{\text{cc}})$$
We name the imputation algorithm as SISNN (see \Cref{alg: sisnn} in \Cref{app:misnn general}).
\end{rem}

\vspace{-0.3cm}
\begin{algorithm}[!htb] 
\caption{\small Multiple Imputation via Semi-parametric Neural Network (MISNN)}
\textbf{Input:} Incomplete data $\D$, number of imputation $M$

\begin{algorithmic}[1]
\STATE Fit a regularized regression $\D_{\text{cc}, 1}\sim \D_{\text{cc},-1}$, with the penalty function $P$, by
$$(\widehat{\bm\alpha},\hat\alpha_0):=\underset{(a,a_0)}{\operatorname{argmin}} \frac{1}{2}\|\D_{\text{cc}, 1}- \D_{\text{cc},-1}\bm a-a_0\|^2 + P(\bm a).$$

\STATE Obtain the active set $ \mathcal{S}:=\{i:\hat\alpha_i\neq 0\}$ and split $\D_{-1}$ into sub-matrices $\X=[\D_{-1}]_\S$ and $\T=\D_{-1}\backslash\X$.

\STATE Given the training data $\left\{\T_{\text{cc}},\D_{\text{cc}, 1},\X_{\text{cc}}\right\}$, train neural networks to learn 
$$\eta_{D}(\T):=\E(\D_{1}|\T), \eta_X(\T):=\E(\X|\T)$$

\STATE Apply standard maximum likelihood technique onto
$$\D_{\text{cc}, 1}-\E(\D_{\text{cc}, 1}|\T_{\text{cc}})= \left(\X_{\text{cc}}-\E(\X_{\text{cc}}|\T_{\text{cc}})\right)\bm\beta+\bm\epsilon$$
where $\bm\epsilon \sim \mathcal{N}(0,\sigma^2)$ and approximate the distribution 
$
\rho_2\left(\bm\beta,\sigma \Big| \D_{\text{obs}, 1},\X_{\text{obs}},\T_{\text{obs}}\right)
$

\FOR{\texttt{$m \in \{1,\dots,M\}$}}
  
\STATE  Randomly draw $ \hat{\bet}^{(m)},\hat{\sigma}^{(m)} $ from the conditional distribution $\rho_2\left(\bm\beta,\sigma \Big| \D_{\text{cc}, 1},\X_{\text{cc}},\T_{\text{cc}}\right)$. 

Subsequently, impute $ \D_{\text{ic}, 1} $ with $ \widehat\D_{\text{ic}, 1}^{(m)} $ by drawing randomly from the predictive distribution $ \rho_1\left(\D_{\text{ic}, 1} | \X_{\text{ic}},\T_{\text{ic}}, \hat{\boldsymbol{\beta}}^{(m)}, \hat{\boldsymbol{\sigma}}^{(m)^2}\right)$

\ENDFOR
\end{algorithmic}
\label{alg: misnn} 
\end{algorithm} 

\vspace{-0.4cm}
\subsection{Sampling from Posterior and Predictive Distributions}
\label{app:MISNN GPLM}
\vspace{-0.2cm}

To conduct multiple imputation in MISNN, we need to sample the parameters from the posterior distribution $\rho_2\left(\bm\beta,\sigma^2 \Big| \D_{\text{obs}, 1},\X_{\text{obs}},\T_{\text{obs}}\right)$ and the predictive distribution $ \rho_1\left(\D_{\text{miss}, 1} | \X_{\text{miss}},\T_{\text{miss}}, \hat{\boldsymbol{\beta}}^{(m)}, \hat{\boldsymbol{\sigma}}^{(m)^2}\right)$ in MISNN (c.f. \Cref{alg: misnn}). With the partialling out, we fit a linear regression at step (4), 
\small
$$
\D_{\text{obs}, 1}-\E(\D_{\text{obs}, 1}|\T_{\text{obs}})= \left(\X_{\text{obs}}-\E(\X_{\text{obs}}|\T_{\text{obs}})\right)\bm\beta+\bm\epsilon
$$ 
\normalsize
We approximate the posterior distribution of $\bm \beta, \sigma$ using
\small
\begin{align*}
   \rho_2\left(\bm\beta,\sigma^2 \Big| \D_{\text{obs}, 1},\X_{\text{obs}},\T_{\text{obs}}\right)
    = f_1\left(\bm\beta \Big| \D_{\text{obs}, 1},\X_{\text{obs}},\T_{\text{obs}}\right)
   \times f_2\left(\sigma^2\Big|\D_{\text{obs}, 1},\X_{\text{obs}},\T_{\text{obs}}\right)
\end{align*}
\normalsize
Suppose the OLS estimate for $\bm \beta$ and its variance are $\Bar{\bm \beta}$ and $\Sigma_{\bm \beta}$, respectively. We can approximate the distribution of $\bm \beta$ by a normal distribution:
\small
\begin{align*}
    f_1\left(\bm\beta \Big| \D_{\text{obs}, 1},\X_{\text{obs}},\T_{\text{obs}}\right) \sim \mathcal{N}\left(\Bar{\bm \beta},\Sigma_{\bm \beta} \right)
\end{align*}
\normalsize
where the parameters are defined as:
\small
$$\Bar{\bm\beta}=\text{argmin}_b \|\D_{\text{obs}, 1}-\eta_{D}(\T_{\text{obs}})-[\X_{\text{obs}}- \eta_X(\T_{\text{obs}})]\bm b\|^2$$
$$\Sigma_{\bm \beta} = \Bar{\sigma}^2 \left( (\X_{\text{obs}}- \eta_X(\T_{\text{obs}})^{\top}(\X_{\text{obs}}- \eta_X(\T_{\text{obs}}))\right)^{-1}$$
\normalsize
Here $\Bar\sigma^2$ can be estimated as the mean of squared residuals: 
\small
\begin{align*}
f_2\left(\sigma^2\Big|\D_{\text{obs}, 1},\X_{\text{obs}},\T_{\text{obs}}\right)
    =\left\| \D_{\text{obs}, 1}-\eta_{D}(\T_{\text{obs}})-( \X_{\text{obs}}- \eta_X(\T_{\text{obs}}))\Bar{\bm\beta} \right\|^2/n_{\text{obs}}  
\end{align*}
\normalsize
As for drawing from the predictive distribution, we calculate $\hat\sigma^{(m)}$ from $f_2$ (with $\Bar\beta$ substituted by $\bm{\hat\beta}^{(m)}$).
At last, we can draw $\widehat\D_{\text{miss}, 1}^{(m)}$ from
\small
\begin{align*}
    \rho_1\left(\D_{\text{miss}, 1} | \X_{\text{miss}},\T_{\text{miss}}, \hat{\boldsymbol{\beta}}^{(m)}, \hat{\boldsymbol{\sigma}}^{(m)^2}\right) 
   = \eta_{D}(\T_{\text{miss}})+(\X_{\text{miss}}-\eta_X(\T_{\text{miss}}))\hat{\boldsymbol{\beta}}^{(m)}
+\mathcal{N}(0,\hat{\boldsymbol{\sigma}}^{(m)^2})
\end{align*}
\normalsize

\vspace{-0.4cm}
\subsection{Flexibility of MISNN Framework}
\vspace{-0.2cm}

Again, we highlight that the framework of MISNN is flexible in two folds: It can incorporate arbitrary feature selection method and arbitrary neural network models during imputation.

MISNN can incorporate an arbitrary feature selection method. Here, we adopt Lasso to select features $\X=\D_\mathcal{S}$ and $\T=\D_{-1} \setminus \D_\mathcal{S}$, where $\mathcal{S} = \{i>0:\hat\alpha_i\neq 0\}$ comes from the non-zero part of lasso estimate
$$(\hat{\bm\alpha},\hat\alpha_0)=\text{argmin}_{\bm{a},a_0}\frac{1}{2}\|\D_{\text{cc},1}-\D_{\text{cc},-1}\bm{a}-a_0\|_2^2+\lambda\|\bm{a}\|_1$$


MISNN works compatibly with all types of networks. Especially, when equipped with over-parameterized neural networks, MISNN can borrow the results from DebiNet \cite[Theorem 1\&2]{xu2020debinet} to claim $\sqrt{n}$-consistency and exponentially fast convergence.



In practice, MISNN can work with a much richer class of neural networks than those theoretically supported in the neural tangent kernel regime \cite{du2019gradient,allen2019convergence}. This includes the under-parameterized, moderately wide and deep neural networks. Empirical experiments shows that PLM learned by such neural networks exhibit strong prediction accuracy as well as post-selection inference (see \Cref{table: real-data}).

\vspace{-0.4cm}
\subsection{Other Properties of MISNN}
\vspace{-0.2cm}

Here we discuss some properties that MISNN enjoys, besides the flexibility of the framework, the consistent estimation of $\bm\beta$ and the fast training of PLM aforementioned.  Numerical evidence can be found in \Cref{sec:experiment}.

\textbf{Trainability: } MISNN can be trained by existing optimizers in an efficient manner, in comparison to Bayesian Lasso (which may require expensive burn-in period, see \Cref{table:burn-in}), boostrap methods (e.g. DURR, which needs many bootstrapped subsamples to be accurate) or MICE (which fits each feature iteratively and may be slow in high dimension).

\textbf{Robustness: } Empirically, MISNN is robust to hyper-parameter tuning (e.g. the width of hidden layers does not affect the performance much). From the data perspective, in high feature dimension and high missing rate (e.g. when compared to DURR, IURR and GAIN), MISNN still works reasonably well.

\vspace{-0.4cm}
\subsection{MISNN for General Missing Patterns}
\label{sec: misnn general}
\vspace{-0.2cm}

The imputation procedure can be naturally extended to the case of general missing patterns, in which the pseudo code is provided in \Cref{alg: misnn general} in the \Cref{app:misnn general}. Suppose the first $K$ columns are missing in $\D$, denoted as $\D_{\text{full}, [K]}$ and the $k$-th column is denoted by $\D_{\text{full}, k}$. The set $-[K]$ represents all other columns except those in $[K]$. Similar to the case of single  column missingness, to construct a partial linear model, we need to partition the data into $\X$ and $\T$. We fit regularized linear regression for each of the $K$ columns that have missing values and obtained $K$ active sets. Then we propose to use either intersection or union to combine the sets into a single one, which will be treated as $\X$. To estimate the parameters $\beta$, during each imputation, for the $k$-th column, we consider an OLS model that uses $\D_{\text{full}, [K]}$ as regressors and the $k$-th column as response. Maximum likelihood techniques are adopted to generate regression coefficients ${\bet}_k$. 


We remark that other proper feature selection methods and set-merging rules can be adopted to replace what we use. It's also possible that we use an iterative approach, following the idea of MICE, to conduct column-wise imputation. Generalization to the case of discrete missing values can be realized with the help of GPLM, which is similar to the discussion in \Cref{sec: gplm}.

\vspace{-0.4cm}
\section{Numerical Results}
\label{sec:experiment}
\vspace{-0.2cm}


We compared MISNN with other state-of-the-art methods on various synthetic and real-world datasets. To establish baselines, we included complete data analysis, complete case analysis, and column mean imputation. We also evaluated two MI methods that incorporate regularized linear models for feature selection in high-dimensional settings: MICE-DURR and MICE-IURR. Additionally, we included MissForest, a MICE approach that uses random forest as the regression model, as well as GAIN, a deep-learning-based imputation method, and two matrix completion methods: SoftImpute and MMMF. More details about our experimental setup and results can be found in \Cref{app:experiment}.


\begin{table*}[!htb]
\vspace{-0.5cm}
    \small
	\centering
\hspace*{-0.4cm}
\begin{tabular}{|C{4cm}|C{0.7cm}|C{1.2cm}|C{1.35cm}|C{1.3cm}|C{1.3cm}|C{1cm}|C{1cm}|}
	\hline Method &Style& Bias & Imp MSE & Coverage & Seconds&SE&SD\\
	\hline 
	Complete Data &- &\textbf{0.0027} &- &\textbf{0.954} & - &0.1126 &0.1150 \\
	Complete Case &- &0.1333 &- &0.854 & - &0.1556 &0.1605\\
	Mean-Impute &SI &0.1508 &12.6215 &\textbf{0.994} &\textbf{0.005} &0.3268 &0.1933 \\
	\hline
	MISNN-wide (Lasso) &MI& \textbf{-0.0184} & \textbf{4.2382} & \textbf{0.902} & \textbf{0.324} & 0.1438 & 0.1713 \\
	MISNN-wide (ElasticNet) &MI& \textbf{-0.0134} & \textbf{4.2191} & \textbf{0.924} & \textbf{0.286} & 0.1431 & 0.1641 \\
	MISNN-narrow (Lasso) &MI& \textbf{-0.0251} & 6.2666 & \textbf{0.944} & \textbf{0.370} & 0.1816  & 0.1755 \\
	MISNN-narrow (ElasticNet) &MI& \textbf{-0.0246} & 6.2550 & \textbf{0.956} & \textbf{0.344} & 0.1818  & 0.1647 \\
	MICE-DURR (Lasso) &MI&0.1815 &12.6704 &\textbf{0.978} &1.266 &0.2275 &0.1196 \\
	MICE-DURR (ElasticNet) &MI&0.1314 &10.8060 &\textbf{0.990} &0.633 &0.2241 &0.1219 \\
	MICE-IURR (Lasso) &MI&0.2527 &15.7803 &0.886 &1.483 &0.2136 &0.1150 \\
	MICE-IURR (ElasticNet) &MI&0.2445 &15.3266 &0.892 &0.566 &0.2153 &0.1399 \\
	MissForest &MI&0.0579 &9.6174 &\textbf{0.962} &69.948 &0.2851 &0.2609 \\
    GAIN &SI&0.7578 &27.3505 &0.289 &14.812 &0.2869 &0.4314 \\
	SoftImpute&SI&-0.1432 &\textbf{4.6206} &0.842 &\textbf{0.019} &0.1804 &0.2005\\
	MMMF&SI&-0.1239 &\textbf{4.0956} &0.782 &3.385 &0.1491 &0.1869\\
	\hline
\end{tabular}
\vspace*{0.2cm}
\caption{Multi-feature missing pattern in synthetic data over 500 Monte Carlo datasets. Bias: mean bias $\hat\beta_1-\beta_1$; Imp MSE: $\|\widehat\D_{\text{miss},1:3}-\D_{\text{miss},1:3}\|^2/n_{\text{miss}}$; Coverage: coverage probability of the 95\% confidence interval for $\beta_1$; Seconds: wall-clock imputation time; SE: mean standard error of $ \hat{\beta}_{1}$; SD: Monte Carlo standard deviation of $ \hat{\beta}_{1}$. Model settings are in \Cref{sec:multi-column missing} and data generation is left in \Cref{sec:model multiple column}.}
\vspace{-0.9cm}
\label{table: multi-col}
\end{table*}



In addition to imputation accuracy, we evaluate the performance of imputation models in statistical inference that are based on imputed datasets. In all the experiments, we specify a set of predictors and a response in the data matrix $\D = (\Z, y)$. A linear regression 
$
\hat{y} = \Z\hat{\bm\theta}
$
is fitted using imputed dataset to predict $y$ and we record the regression parameters $\hat{\bm\theta}$. In synthetic datasets, we have access to the ground truth $\bm\theta$, so we focus on inference performance. In real data analysis, we lose access to the true $\bm\theta$ and focus on the prediction error instead.

\vspace{-0.4cm}
\subsection{Viewpoint of Statistical Inference}
\vspace{-0.2cm}
\label{sec:multi-column missing}

In terms of the statistical inference, we consider four statistical quantities: bias of $\hat{\bm\theta}$, coverage rate of the 95\% confidence interval (CR) for ${\bm\theta}$, mean standard error (SE) for $\hat{\bm\theta}$ and Monte Carlo standard deviation (SD) of $\hat{\bm\theta}$. Imputation mean squared error (MSE) is also compared. We study the performance of MISNN under general missing patterns, in which multiple columns (features) in the dataset can contain missing values. We adopt a similar experiment setting to that in \cite{deng2016multiple} and evaluate performance over 500 Monte Carlo datasets. A detailed experiment description can be found in Appendix \ref{app:experiment}.



\begin{table*}[!htb]
\vspace{-0.4cm}
	\centering
\begin{tabular}{|C{4cm}|C{1cm}|C{1.3cm}|C{1.35cm}|C{1.3cm}|C{1.3cm}|C{1.5cm}|}
	\hline Method &Style& Estimator & Imp MSE & Seconds&SE &Pred MSE\\ 
	\hline 
	Complete Data &- &\textbf{0.0532} &- &- &0.0676 &\textbf{0.8695}\\
	Complete Case &- &0.1278 &- &- &0.1392  &1.3376\\
	Mean-Impute &SI &-0.0374 &1.3464 &\textbf{0.006} &0.0686 &0.8938\\
	\hline
	MISNN (Lasso) &MI& \textbf{0.0545} & 0.6620 & \textbf{1.501} &  0.0681  &\textbf{0.8780}\\
	MISNN (ElasticNet) &MI& \textbf{0.0521} & \textbf{0.5140} & \textbf{0.861} & 0.0716  &\textbf{0.8789}\\
	MICE-DURR (Lasso) &MI&\textbf{0.0504} &1.8256 &3.946 &0.0508  &\textbf{0.8755}\\
	MICE-DURR (ElasticNet) &MI&0.0426 &1.6998 &2.709 &0.0552 &0.8817\\
	MICE-IURR (Lasso) &MI& 0.0474 &2.0404 &4.093  &0.0476  & \textbf{0.8747}\\
	MICE-IURR (ElasticNet) &MI&0.0318 & 2.0219 &2.620 &0.0484  &0.8803\\
	GAIN &SI&0.0304 &0.9902 &67.432 &0.0504  &\textbf{0.8749}\\
	SoftImpute&SI& \textbf{0.0533} &0.6667 & \textbf{0.0344} &0.0763  &0.8808\\
	MMMF&SI&0.0833  &\textbf{0.3051} &5.0261 &0.0838  &\textbf{0.8755}\\
	\hline
\end{tabular}
\caption{Multi-feature missing pattern in ADNI dataset over 100 repeats. Estimator: estimated $\hat\beta_1$ through OLS using first 5 features as regressors; Imp MSE: imputation mean squared error $\|\widehat\D_{\text{miss},1:3}-\D_{\text{miss},1:3}\|^2/n_{\text{miss}}$; Seconds: wall-clock imputation time; SE: mean standard error of $ \hat{\beta}_{1}$; Pred MSE: mean squared error between $\A\bm{\hat\theta}$ and $\y$. Model settings are in \Cref{sec:real missing} and data generation is left in \Cref{sec:model real data}. MissForest is too slow (more than 5 min per dataset) to be considered.}
\vspace{-0.9cm}
\label{table: real-data}
\end{table*}

Potentially, one can combine MICE with MISNN for single-column missingness as well. Nevertheless, we avoid doing so by proposing \Cref{alg: misnn general} in \Cref{app:misnn general}, which deals with the general missing patterns differently, in a parallel computing fashion. During the experiments, we use different network structures at step (3) of Algorithm \ref{alg: misnn general}: MISNN-wide uses two hidden layers with width 500, each followed by ReLU activation, a Batch Normalization layer \cite{ioffe2015batch} and a Dropout layer \cite{srivastava2014dropout} at rate 0.1. The neural networks in MISNN-narrow are the same as in MISNN-wide, except the hidden layers have width 50 instead.

The results are summarized in Table \ref{table: multi-col}. We highlight that all MISNN give the smallest estimation bias compared with the rest of imputation methods. MISNN also achieves satisfying imputation MSE, statistical coverage and computation speed. In comparison, two matrix completion methods achieve comparable imputation MSE, but their coverage is much worse than MI methods.

It is interesting to note that MISNN-wide tends to have smaller imputation MSE and estimation bias than MISNN-narrow. However, the coverage of the former is not as good as the latter, mainly due to the small SE. We suggest that in practice, if the accuracy of imputation or the parameter estimation is of main interest, MISNN with wide hidden layers should be adopted. If the statistical inference on parameters of interest is emphasized, then MISNN should be equipped with narrow hidden layers.

\vspace{-0.4cm}
\subsection{Viewpoint of Prediction}
\vspace{-0.2cm}
\label{sec:real missing}

We applied MISNN to the Alzheimer's Disease Neuroimaging Initiative (ADNI) gene dataset \footnote{The complete ADNI Acknowledgement is available at \url{http://adni.loni.usc.edu/wp-content/uploads/how_to_apply/ADNI_Acknowledgement_List.pdf}.}, which includes over 19k genomic features for 649 patients and a response, VBM right hippocampal volume, ranging between [0.4,0.6]. We selected the top 1000 features with the largest correlations with the response, and focused on the linear analysis model between the response and the top 5 features. Since we did not have access to the true coefficients in the linear model, we studied the difference between the estimated coefficients from complete data analysis and the ones from imputed datasets. We artificially generated missing values under MAR in the top 3 features that had the largest correlations with the response, with a missing rate of approximately $65\%$. We used MISNN, containing a single hidden layer with width 500 and a Batch Normalization layer, and fit a linear regression between the response $\y$ and the top five features ${\D_1 \sim \D_5}$ for downstream prediction.

Our results, summarized in Table \ref{table: real-data}, show that MISNN achieved small imputation and prediction MSEs in a computationally efficient manner, particularly when compared to other MI methods. Additionally, the estimators by MISNN (as well as SoftImpute) were closest to the gold criterion from complete data analysis. Further experiment details can be found in Appendix \ref{app:experiment}.



\vspace{-0.4cm}
\section{Discussion}
\vspace{-0.2cm}

In this work, we propose MISNN, a novel deep-learning based method for multiple imputation of missing values in tabular / matrix data. We demonstrate that MISNN can flexibly work with any feature selection and any neural network architecture. MISNN can be trained with off-the-shelf optimizers at high computation speed, providing interpretability for the imputation model, as well as being robust against data dimension and missing rate. Various experiments with synthetic and real-world datasets illustrate that MISNN significantly outperforms state-of-the-art imputation models.

While MISNN works for a wide range of analysis models, we have only discussed the case for continuous missing values using the partialling out. We can easily extend MISNN to discrete missing value problems by considering the generalized partially linear models (GPLM, see \Cref{app:MISNN GPLM} for details). However, the partialling out technique generally renders invalid for GPLM. Therefore, iterative methods including the backfitting, which can be slow, may be required to learn MISNN. 

\vspace{-0.4cm}
\section{Acknowledgement}
\vspace{-0.2cm}
This work was supported in part by National Institutes of Health grant, R01GM124111. The content is solely the responsibility of the authors and does not necessarily represent the official views of the National Institutes of Health.

\vspace{-0.4cm}
\bibliography{ref}
\vspace{-0.2cm}

\clearpage
\appendix

\section{Analysis model and Rubin's rule}
\label{rubin's rule}

In this section, we explain the purpose and theoretical foundation of Rubin's rule. Suppose the estimand $\boldsymbol{\theta}$ is what we are eventually interested in and could be calculated if we observe the complete data set. After collecting the $M$ imputed datasets, we utilize the Rubin's rule to infer $\bm{\bar\theta}$ with careful quantification of the model uncertainty. Detailed implementation of Rubin's rule is described in \Cref{alg: Rubin}.

\begin{algorithm}[!htpb]
\caption{Rubin's rule for confidence interval}
\textbf{Input:} $M$ imputed datasets, significance level $s$\\

\begin{algorithmic}
\STATE (1) Calculate $\boldsymbol{\hat{\theta}}^{(m)}$ and $\text{SE}\left(\boldsymbol{\hat{\theta}}^{(m)}\right)$ from the $m$-th imputed dataset according to the analysis model

\STATE (2) Compute the pooled mean by
$\boldsymbol{\Bar{\theta}} = \sum_{m=1}^M \boldsymbol{\hat{\theta}}^{(m)} $
\STATE (3)  Compute the pooled variance by
\begin{align*}
    \text{Var}_{\text{within}} &= \frac{\sum_{m=1}^M \text{SE}\left(\boldsymbol{\hat{\theta}}^{(m)}\right)^2}{M}  \\
\text{Var}_{\text{between}} &= \frac{\sum_{m=1}^M (\boldsymbol{\hat{\theta}}^{(m)} - \boldsymbol{\Bar{\theta}})(\boldsymbol{\hat{\theta}}^{(m)} - \boldsymbol{\Bar{\theta}})^{\top}}{M-1}  \\
  \text{Var}_{\text{total}} & =  \text{Var}_{\text{within}} + \left(1+\frac{1}{M} \right) \text{Var}_{\text{between}} 
\end{align*}

\STATE (4) Construct the $(1-s)$ confidence interval with endpoints: with $\Phi$ being the cumulative density function of standard normal,
\begin{align*}
    \boldsymbol{\Bar{\theta}}\pm\Phi^{-1}\left(1-\frac{s}{2}\right)\sqrt{\text{Var}_{\text{total}}}
\end{align*}
\end{algorithmic}
\label{alg: Rubin}
\end{algorithm}

\begin{rem}
Given the pooled mean and pooled variance, there are other ways to construct the confidence interval. For example, one can replace the Gaussian quantile $\Phi^{-1}\left(1-\frac{s}{2}\right)$ by the $t$-distribution quantile $t_{\textup{df},1-\frac{s}{2}}$, where common choices for degree of freedom (\textup{df}) is $\textup{df}_{\text{old}}$ \cite{barnard1999miscellanea} or $\textup{df}_{\text{adjusted}}$ \cite{rubin2004multiple}.
\end{rem}

Here we provide a explanation for Rubin's rule. By the law of total expectation:
\begin{align*}
    \mathbb{E}(\boldsymbol{\theta}|\D_{\text{obs}})=\mathbb{E}(\mathbb{E}[\boldsymbol{\theta}|\D_{\text{obs}},\D_{\text{miss}}]|\D_{\text{obs}})
\end{align*}
This equation motivates to adopt Rubin's rule when combining the results of multiple imputations. Suppose $\boldsymbol{\hat{\theta}}^{(m)}$ is the estimate of the $m$-th imputation, then the pooled mean equals
\begin{align*}
\boldsymbol{\Bar{\theta}} = \sum_{m=1}^M \boldsymbol{\hat{\theta}}^{(m)} 
\end{align*}
The posterior variance of $\boldsymbol{\theta}$ over observed data comes from two sources, by the law of total variance:
\begin{align*}
    \text{Var}(\boldsymbol{\theta}|\D_{\text{obs}}) &= \mathbb{E}[\text{Var}(\boldsymbol{\theta}|\D_{\text{obs}},\D_{\text{miss}})|\D_{\text{obs}}] \\ &+ \text{Var}[\mathbb{E}(\boldsymbol{\theta}|\D_{\text{obs}},\D_{\text{miss}})|\D_{\text{obs}}]
\end{align*}
The first part is the mean of posterior variance of $\boldsymbol{\theta}$ over imputed datasets, named as the `within variance'. The second part is the variance between posterior means of $\boldsymbol{\theta}$, named as the `between variance'. Given the $M$ imputed datasets, the within variance could be estimated by 
\begin{align*}
    \text{Var}_{\text{within}} &= \frac{\sum_{m=1}^M \text{Var}\left(\boldsymbol{\hat{\theta}}^{(m)}\right)}{M}
\end{align*}
The between variance could be estimated by 
\begin{align*}
     \text{Var}_{\text{between}} &= \frac{\sum_{m=1}^M (\boldsymbol{\hat{\theta}}^{(m)} - \boldsymbol{\Bar{\theta}})(\boldsymbol{\hat{\theta}}^{(m)} - \boldsymbol{\Bar{\theta}})^{\top}}{M-1}
\end{align*}
where $\boldsymbol{\Bar{\theta}}$ is calculated as above. 

We emphasize that directly summing $\text{Var}_{\text{within}}$ and $\text{Var}_{\text{between}}$ to derive the total variance is not correct and under-estimates the variance. Because $\boldsymbol{\Bar{\theta}}$ is estimated with a finite number of $M$ datasets, instead of with $M\to \infty$. The difference between the finite estimation and the infinite estimation can be characterized by $\frac{1}{M}\text{Var}_{\text{between}}$\cite{rubin2004multiple}. Therefore all in all, the posterior variance of $\boldsymbol{\theta}$ is estimated by Rubin's rule as
\begin{align*}
     \text{Var}_{\text{total}} & =  \text{Var}_{\text{within}} + \left(1+\frac{1}{M} \right) \text{Var}_{\text{between}} 
\end{align*}




\section{Non-Gaussian Missing Values}
\label{sec: gplm}
When the missing values are not Gaussian, we fit a generalized partially linear model (GPLM) at step (3) of \Cref{alg: misnn} and the algorithm differs from \Cref{alg: misnn} from here.
\begin{align*}
\E(\D_1|\X,\T)&=h^{-1}\left(\alpha_0+\D_{\text{full},\S}\bm\beta+\overline{\D_{\text{full},\S}}\bm\gamma\right)
\\
&=h^{-1}\left(\alpha_0+\X\bm\beta+f(\T)\right)
\end{align*}
For instance, when the missing values are binary, we fit the logistic partially linear model with $h$ being the logit function. 

However, we do not estimate the conditional expectations $\E(\D_1|\T)$ and $\E(\X|\T)$ as the partialling out technique does not apply to GPLM, due the non-linearity of $h$. Instead, methods such as the backfitting are used to approximate the posterior distribution of $\bm\beta$ and the function $\hat f$ in the GPLM.

Treating any fitting method as a black box to learn GPLM, we give two examples of multiple imputation when
\begin{itemize}
    \item $\D_1$ follows Bernoulli distribution, then $h(z)=\log(\frac{z}{1-z})$. We draw $\hat{\bm\beta}^{(m)}$ from the posterior distribution of $\bm\beta$ and further draw the binary imputed values from
\begin{align*}
\widehat\D_{i, 1}^{(m)}&\sim\text{Bernoulli}(\pi_i^{(m)})
\\
\pi_i^{(m)}&=\frac{1}{1+\exp(-\X_i\hat{\boldsymbol{\beta}}^{(m)}-\hat f(\T_i))}
\end{align*}

\item $\D_1$ follows Poisson distribution, then $h(z)=\log(z)$. We draw $\hat{\bm\beta}^{(m)}$ from the posterior distribution of $\bm\beta$ and further draw the positive and discrete imputed values from
\begin{align*}
\widehat\D_{i, 1}^{(m)}&\sim\text{Poisson}(\mu_i^{(m)})
\\
\mu_i^{(m)}&=\exp(\X_i\hat{\boldsymbol{\beta}}^{(m)}+\hat f(\T_i))
\end{align*}
\end{itemize}

The Rubin's rule then applies in the same way as \Cref{alg: Rubin}.

\section{Experiments Details}
\label{app:experiment}

Before going to the details of the experiments, we would like to provide a brief introduction of other MI methods that adopts feature selection.

\paragraph{MI via Bayesian Lasso}

Bayesian Lasso \cite{park2008bayesian} formulates a hierarchical Bayesian model for Lasso, by assigning a double-exponential (Laplacian) prior on the regression coefficients $\bm \alpha$:
\begin{equation}
\label{blasso}
\rho\left(\boldsymbol{\alpha} \mid \sigma^{2}, \lambda, \rho\right)=\prod_{j=1}^{p-1}\frac{\lambda}{2 \sqrt{\sigma^{2}}} \exp \frac{-\lambda\left|\alpha_{j}\right|}{\sqrt{\sigma^{2}}} 
\end{equation}
Here $a,b,r,s$ are pre-specified hyperparameters that govern the prior distribution of variance $\sigma^2$ and penalty parameter $\lambda$:  $\sigma^{2} \sim \operatorname{Inverse-Gamma}(a, b)$ and $\lambda \sim \operatorname{Gamma}(r, s)$. The sampling procedure is in general conducted through Markov Chain Monte Carlo (MCMC). However, in the high-dimensional cases, the sampling procedure is extremely time-consuming and renders Bayesian Lasso accurate yet impractical.

\begin{rem}[Bayesian Lasso must burn-in]
A large number of burn-in iterations during the sampling process are necessary in order to get satisfactory results from Bayesian Lasso. As a result, the computational cost can be high.

\begin{table}[!htb]
	\centering
\begin{tabular}{|l|l|l|}
	\hline Burn-in iterations & Bias &Imp MSE\\
	\hline 50 & -0.7752 & 2.9925 \\
	\hline 500 & -0.4774 & 1.6328 \\
	\hline 5000 & -0.1986 & 0.9249 \\
	\hline 10000 & -0.1979 & 0.9244 \\
	\hline
\end{tabular}
\caption{Performance of Bayesian Lasso for different burn-in periods in the setting of \Cref{table: single-col}.}
\label{table:burn-in}
\vspace{-0.3cm}
\end{table}
\end{rem}

Two alternative MI approaches: DURR and IURR are proposed \cite{zhao2016multiple,deng2016multiple}, both of which are more computationally efficient than Bayesian Lasso and applicable to other regularization beyond Lasso.

\paragraph{MI via direct use of regularized regression (DURR)}\quad
In the $m$-th imputation, DURR $\langle1\rangle$ generates bootstrap dataset $\D^{(m)}$ of size $n$ by sampling with replacement from original dataset $\D$, $\langle2\rangle$ uses regularized regression to obtain estimate $\bm \alpha^{(m)}$, $\langle3\rangle$ imputes missing values by the predictive distribution with $\bm \alpha^{(m)}$. 
Noticeably, DURR is the only method that uses bootstrap to approximate the posterior distribution of $\bm\alpha$.

\paragraph{MI via indirect use of regularized regression (IURR)}
In the $m$-th imputation, IURR $\langle 1\rangle$ fits regularized regression and identifies the active set $\S$, $\langle2\rangle$ uses maximum likelihood to approximate posterior distribution of $\bm\alpha$, $\langle3\rangle$ imputes missing values by the predictive distribution with $\bm\alpha^{(m)}$. The idea of IURR is to only infer on the selected (important) entries and ignore the others. The first two steps combined is known as OLS post-Lasso \cite{belloni2013least} and differs from MISNN as we use PLM, which additionally leverages the information in unselected features.


\subsection{Synthetic missing data: single missing column}
\label{sec:model single column}

In this subsection, we compare the performance of MISNN with other state-of-the-art impuation methods on the synthetic data of univariate missing patterns. We study similar settings as that in \cite{zhao2016multiple}. In all simulations, we fix the sample size $n=100$ and number of features $p=1000$. Each simulated data set includes the label $\y$, and the set of predictors $\D\in\mathbb{R}^{n\times p}$, where $ \D_{1} $ contains missing values. $\left(\D_{2}, \ldots, \D_{p}\right)$ are generated from a multivariate normal distribution with mean 0 and a first-order autoregressive covariance matrix with autocorrelation $0.5$.  $\D_{1} $ is generated from a normal distribution with variance 1 and mean $\alpha_{0}+\mathbf{D}_{S}{\bm\alpha}, $ where $\mathbf{D}_{S}=\left\{\mathbf{D}_{2}, \ldots, \mathbf{D}_{11}, \mathbf{D}_{50}, \ldots, \mathbf{D}_{59}\right\}$ and $\alpha_j = \sqrt{0.2}$. The label $\mathbf{y} $ is generated from a normal distribution with mean $\beta_{0}+\beta_{1} \mathbf{D}_{1}+\beta_{2} \mathbf{D}_{2}+\beta_{3} \mathbf{D}_{3}\left(\beta_{j}=1,\A=[\D_1,\D_2,\D_3]\right) $ and variance 1. 

To generate data that are MAR, the missing value indicator follows $\text{logit}[\operatorname{Pr}(\mathbf{D}_{1}$ $\text{ is missing} | \mathbf{D}_{-1}, \mathbf{y})]=3-0.1 \mathbf{D}_{2}+3 \mathbf{D}_{3}-2 \mathbf{y}, $ hence approximately
$ 50 \% $ of $ \mathbf{D}_{1} $ is missing. We summarize the performances of different imputation methods in Table \ref{table: single-col}. Clearly, MISNN is robust to such high dimension and high missing rate: MISNN is significantly faster than other methods, e.g. Bayesian Lasso, that demonstrate reasonable coverage rate and small bias. Although matrix completion methods and Lasso exhibit the smallest imputation MSE, these SI methods lead to improper inference with biased estimate of $\beta_1$ and low coverage. Of note, GAIN shows the worst performance because it only works for MCAR mechanism.

Throughout we use same Lasso penalty ($\lambda=0.1$) for all the methods which require a feature selection step. For multiple imputation methods, 30 imputed datasets are generated. We use R to implement Bayesian Lasso (R package \texttt{monomvn} \cite{gramacy2010monomvn}) with 10000 burn-in iterations and $(a,b,r,s) = (0.1,0.1,0.01,0.01)$ in the prior model (\ref{blasso}) and use tensorflow to implement GAIN \cite{yoon2018gain}. We use python/pythorch to implement others methods.   

\textbf{Details of MISNN:} Here we adopted two-layer fully connected neural network as the structure of MISNN. The width of the hidden layer is 500. Activation function is ReLU and a batch normalization is applied. We use Adam optimizer and the learning rate is $10^{-3}$. An early stopping mechanism is applied and the patience is $1$.

\begin{table*}[!htb]
	\centering
\begin{tabular}{|C{2.5cm}|C{1cm}|C{1.4cm}|C{1.4 cm}|C{1.4cm}|C{1.4cm}|C{1.4cm}|C{1.4cm}|}
	\hline Method &Style& Bias & Imp MSE & Coverage & Seconds&SE&SD\\
	\hline 
	Complete Data &-&\textbf{-0.0057} & - & \textbf{0.951} & - & 0.1412 &0.1419 \\
	Complete Case &-& 0.2052 & - & 0.754 & - & 0.1823 & 0.1977\\
	Mean-Impute &SI& 0.8086 & 2.9364& 0.021 & \textbf{0.014} & 0.2145 & 0.1833 \\
	\hline
	MISNN-Lasso &MI& \textbf{0.0894} & \textbf{0.8431} & \textbf{0.938} & \textbf{0.114} & 0.1844 & 0.1690 \\
	Bayesian Lasso &MI& -0.1979 & \textbf{0.9244} & \textbf{0.840} & 82.72 & 0.2035 & 0.2708 \\
	DURR-Lasso &MI& 0.4244 & 1.5518 & 0.318 & \textbf{0.022} &0.1798 &0.1305 \\
	IURR-Lasso &MI& 0.4542 & 1.7734 & 0.146 &  \textbf{0.038} & 0.1551 &0.1099 \\
	GAIN &SI& 0.6277 & 3.7016 & 0.430 &  89.75 & 0.1500 &0.4970 \\
	SoftImpute&SI& -0.4199& \textbf{0.4306}& 0.320& \textbf{0.043}& 0.1599& 0.2369\\
	MMMF&SI& -0.4384& \textbf{0.4242}& 0.302& 2.070& 0.1668& 0.1999\\
    Lasso &SI& -0.5350 & \textbf{0.4030} & 0.132 &  \textbf{0.032} & 0.1521 &0.2007 \\
	\hline
\end{tabular}
\caption{Single-feature missing pattern in synthetic data over 500 Monte Carlo datasets. Bias: mean bias $\hat\beta_1-\beta_1$; Imp MSE: imputation mean squared error $\|\widehat\D_{\text{miss},1}-\D_{\text{miss},1}\|^2/n_{\text{miss}}$; Coverage: coverage probability of the 95\% confidence interval for $\beta_1$; Seconds: wall-clock imputation time; SE: mean standard error of $ \hat{\beta}_{1}$; SD: Monte Carlo standard deviation of $ \hat{\beta}_{1}$. Data generation and model settings are discussed in \Cref{sec:model single column}.}
\label{table: single-col}
\end{table*}

\subsection{Synthetic missing data: multiple missing columns}
\label{sec:model multiple column}

\textbf{Data generation:} Here we adopted the data generation procedure in \cite{deng2016multiple} where three columns of the dataset contain missing values. Similar with setting in the single-column missing pattern, each simulated dataset has sample size $n=200$ and includes $\y$. Each observation has $p=1000$ features and the first three features contain missing values. We first generate $(\D_4, \dots,\D_p)$ from a multivariate normal distribution with mean 0 and a first-order autoregressive covariance matrix with autocorrelation $0.5$. Given $(\D_4, \dots,\D_p)$, $(\D_1, \D_2,\D_3)$ are generated independently from a normal distribution $\mathcal{N}(0, \D_S\bm\alpha)$, where $\mathbf{D}_{S}=\left\{\mathbf{D}_{2}, \ldots, \mathbf{D}_{11}, \mathbf{D}_{50}, \ldots, \mathbf{D}_{59}\right\}$ and $\alpha_j = \sqrt{0.2}$. The label $\mathbf{y} $ is generated a normal distribution with mean $\beta_{0}+\beta_{1} \mathbf{D}_{1}+\beta_{2} \mathbf{D}_{2}+\beta_{3} \mathbf{D}_{3}+\beta_{4} \mathbf{D}_{4}+\beta_{5} \mathbf{D}_{5}\left(\beta_{j}=1\right) $ and variance 6. 

The missing value is generated in $(\D_1, \D_2,\D_3)$ from logit models. Suppose the missing indicators are $\delta_1,\delta_2,\delta_3$ respectively. Then the logit models are  $\text{logit}(\delta_1=1)=-1-\D_4+2\D_5-\y$, $\text{logit}(\delta_2=1)=-1-\D_4+2\D_{51}-\y$, $\text{logit}(\delta_3=1)=-1-\D_{50}+2\D_{51}-\y$.

At the imputation step, we adopted two feature selection methods: Lasso ($\lambda=0.2$) and ElasticNet ($\lambda=1.0$ with $\ell_1$ penalty ratio 0.5). The penalty is same for all the needed methods. With Adam as the optimizer, the learning rate we adopt for MISNN-wide is $0.01$ and that for MISNN-narrow is $0.001$. We train MISNN-wide for 5 epochs and MISNN-narrow for 15 epochs. We also add a learning rate decay for both networks with decay rate 0.6 for every two steps, with each step being taken after one batch.

\subsection{Data from Alzheimer's Disease study}
\label{sec:model real data}

In this subsection, we provide more details on the experiment of ADNI dataset. The original data contains 649 subjects; each subject includes 19822 features and a continuous response $\y$. We first standardize all the features and response by removing the mean and scaling to unit variance. Then 1000 features which has largest correlation with the response are selected and ranked in a decreasing order. Denote those features as $\A=[\A_1,\dots,\A_{1000}]$. Missing values are generated under MAR in $[\A_1,\A_2,\A_3]$ according to logit models $\text{logit}(\delta_1=1)=-1+2\D_4+\D_5 +2\y$, $\text{logit}(\delta_2=1)=-1+\D_4+2\D_{51}+3\y$, $\text{logit}(\delta_3=1)=-1-\D_{50}+2\D_{51}+\y$. Afterward, the preprocessed dataset $\D$ is obtained by combining feature matrix $\A$ and response $\y$. A pre-specified penalty is used for all the methods which requires feature selection. For Lasso, the penalty $\lambda=0.1$; for ElasticNet, the penalty $\lambda=0.8$ and $\ell_1$ ratio 0.5. The structure of MISNN is the same as that in \cref{sec:model single column}.


\section{MISNN Variants}
\label{app:misnn general}

\begin{algorithm}[!htb] 
\caption{MISNN with Multiple Missing Columns}
\textbf{Input:} Incomplete data $\D$, number of imputation $M$

\begin{algorithmic}
\STATE
\STATE (1)
\FOR{\texttt{$k \in \{1,\dots,K\}$}}
\STATE Fit a regularized regression $\D_{\text{obs}, k}\sim \D_{\text{obs},-[K]}$ by
\begin{align*}
(\widehat{\bm\alpha}_k,\hat\alpha_{0,k}):=\underset{(a,a_0)}{\operatorname{argmin}} \frac{1}{2}\|\D_{\text{obs},k}- \D_{\text{obs},-[K]}\bm a-a_0\|^2 + P(\bm a)
\end{align*}
where $P$ is the penalty function. 
\STATE\STATE
Obtain the active set $ \mathcal{S}_k:=\{i:\hat\alpha_{k,i}\neq 0\}$
\ENDFOR

\STATE\STATE Combine $K$ active sets into a single active set by taking intersection $\mathcal{S}= \cap_{k=1}^K \mathcal{S}_k$ or union $\mathcal{S}= \cup_{k=1}^K \mathcal{S}_k$.

\STATE\STATE (2) Split $\D_{\text{full},-[K]}$ into sub-matrices $\X=[\D_{\text{full},-[K]}]_\S$ and $\T=\D_{\text{full},-[K]}\backslash\X$.

\STATE\STATE (3) Given the training data $\left\{\T_{\text{obs}},\D_{\text{obs}, [K]},\X_{\text{obs}}\right\}$, train neural networks to learn 
$$\eta_{D}(\T):=\E(\D_{\text{full}, [K]}|\T), \eta_X(\T):=\E(\X|\T)$$

\STATE (4) 
\FOR{\texttt{$k \in \{1,\dots,K\}$}}
\STATE Apply standard maximum likelihood technique onto
\begin{align*}
\D_{\text{obs}, k}-\E(\D_{\text{obs}, k}|\T_{\text{obs}})= \left(\X_{\text{obs}}-\E(\X_{\text{obs}}|\T_{\text{obs}})\right)\bm\beta_k+\bm\epsilon
\end{align*}
where $\bm\epsilon \sim \mathcal{N}(0,\sigma_k^2)$ and approximate the posterior distribution 
$$
\rho_2\left(\bm\beta_k,\sigma_k \Big| \D_{\text{obs}, k}-\eta_{D}(\T_{\text{obs}}), \X_{\text{obs}}- \eta_X(\T_{\text{obs}})\right)
$$
\ENDFOR

\STATE (5)\FOR{\texttt{$m \in \{1,\dots,M\}$}}
  \FOR{\texttt{$k \in \{1,\dots,K\}$}}
\STATE  Randomly draw $ \hat{\bet}_k^{(m)},\hat{\sigma}_k^{(m)} $ from posterior distribution $\rho_2\left(\bm\beta_k,\sigma_k \Big| \D_{\text{obs}, k},\X_{\text{obs}},\T_{\text{obs}}\right)$. 

Subsequently, impute $ \D_{\text{miss}, k} $ with $ \widehat\D_{\text{miss}, k}^{(m)} $ by drawing randomly from the predictive distribution $ \rho_1\left(\D_{\text{miss}, k} | \X_{\text{miss}},\T_{\text{miss}}, \hat{\boldsymbol{\beta}}_k^{(m)}, \hat{\boldsymbol{\sigma}}_k^{(m)^2}\right)$
\ENDFOR
\ENDFOR

\end{algorithmic}
\label{alg: misnn general} 
\end{algorithm}

\subsection{MISNN for Single Imputation}
Suppose one only focuses on imputation and prediction tasks, then no sampling from the posterior distribution and the predictive distribution is needed in MISNN. We present the single imputation version of MISNN when dealing with univariate missing patterns, as is shown in \cref{alg: sisnn}. This single imputation algorithm can be trivially extended to cases with discrete missing values or under general missing patterns, as discussed in \Cref{sec: gplm} and \Cref{sec: misnn general}.

\begin{algorithm}[!htb] 
\caption{Single Imputation via Semi-parametric Neural Network (SISNN)}
\textbf{Input:} Incomplete data $\D$, number of imputation $M$

\begin{algorithmic}
\STATE 
\STATE (1) Fit a regularized regression $\D_{\text{obs}, 1}\sim \D_{\text{obs},-1}$ by
$$(\widehat{\bm\alpha},\hat\alpha_0):=\underset{(a,a_0)}{\operatorname{argmin}} \frac{1}{2}\|\D_{\text{obs}, 1}- \D_{\text{obs},-1}\bm a-a_0\|^2 + P(\bm a) $$
where $P$ is the penalty function.

\STATE (2) Obtain the active set $ \mathcal{S}:=\{i:\hat\alpha_i\neq 0\}$ and split $\D_{-1}$ into sub-matrices $\X=[\D_{-1}]_\S$ and $\T=\D_{-1}\backslash\X$.

\STATE (3) Given the training data $\left\{\T_{\text{obs}},\D_{\text{obs}, 1},\X_{\text{obs}}\right\}$, train neural networks to learn 
$$\eta_{D}(\T):=\E(\D_{1}|\T), \eta_X(\T):=\E(\X|\T)$$

\STATE (4) Apply ordinary least squares to derive $\bm{\hat\beta}$ on
$$\D_{\text{obs}, 1}-\E(\D_{\text{obs}, 1}|\T_{\text{obs}})= \left(\X_{\text{obs}}-\E(\X_{\text{obs}}|\T_{\text{obs}})\right)\bm\beta+\bm\epsilon$$

\STATE (5) Impute with $$\D_{\text{miss}, 1}= \left[\X_{\text{miss}}-\eta_X(\T_{\text{miss}})\right]\bm{\hat\beta}+\eta_D(\T_{\text{miss}})$$

\end{algorithmic}
\label{alg: sisnn} 
\end{algorithm}

\end{document}